\documentclass{article}
\pdfoutput=1
\usepackage{graphicx}  
\usepackage{amsmath}   
\usepackage[compress]{cite}
\usepackage{amssymb}   
\usepackage{bm} 
\usepackage{dcolumn}
\usepackage{color}
\usepackage{mathrsfs}
\usepackage{amsfonts}
\usepackage{varioref}
\usepackage[caption=false]{subfig}
\RequirePackage[colorlinks,citecolor=blue,urlcolor=magenta,linkcolor=blue]{hyperref}
\input epsf
\usepackage{accents}

\newcommand*\underdot[1]{%
  \underaccent{\dot}{#1}}
\newcommand*\underddot[1]{%
  \underaccent{\ddot}{#1}}
\labelformat{section}{Section #1} 
\labelformat{subsection}{Section #1} 
\labelformat{subsubsection}{Section #1}
\labelformat{subsubsubsection}{Section #1}
\labelformat{equation}{Eq.~(#1)} 
\labelformat{figure}{Fig.~#1} 
\labelformat{subfigure}{Fig.~\thefigure#1} 
\labelformat{table}{Tab.~#1} 
\labelformat{appendix}{Appendix #1}
\title{Radion induced inflation on non-flat brane and modulus stabilization}

\author{Indrani Banerjee\footnote{tpib@iacs.res.in}~,
Sumanta Chakraborty\footnote{sumantac.physics@gmail.com}~
and Soumitra SenGupta\footnote{tpssg@iacs.res.in}\\
{\small{School of Physical Sciences, Indian Association for the Cultivation of Science, Kolkata-700032, India}}}

\date{ }  
\begin{document}
  
\maketitle 
\begin{abstract}
Warped braneworld models has the potential to provide a plausible resolution to the gauge hierarchy problem but at the cost of introduction of an additional field, namely the inter-brane separation or, the radion field. Providing an appropriate stabilization scheme for the radion field has long been sought for. In this work, we discuss dynamical stabilization of the radion field in the context of early universe cosmology. We show that by allowing the branes to be \emph{non-flat} and \emph{dynamical} a radion potential is naturally generated which can trigger the inflationary phase of the early universe. Interestingly, the kinetic term of the radion during the inflationary epoch exhibits a phantom-like behavior  and the dynamical evolution of the radion field to its stable value marks the exit from the phantom phase as well as the inflationary era simultaneously. We show that the amalgamation and synergy of the radion potential along with its kinetic term can not only trigger the inflationary scenario but also provide a successful exit from the same while maintaining concordance with current observation. 
\end{abstract}
\section{Introduction}
\label{Intro}

Extra dimensional models \cite{Antoniadis:1990ew, Horava:1995qa, Horava:1996ma, PhysRevD.54.R3693, Kakushadze:1998wp, ArkaniHamed:1998rs,  Antoniadis:1998ig, PhysRevD.59.086004, Randall:1999ee, Randall:1999vf, Lykken:1999nb, Kaloper:1999sm}, originally intended to provide grand unification of the known forces also play crucial roles in resolving the gauge-hierarchy problem, arising due to large radiative corrections to the Higgs mass. Even within the realm of extra dimensional physics, there are several alternatives, including the models with large \cite{ArkaniHamed:1998rs, PhysRevD.59.086004,Antoniadis:1998ig} and warped extra dimensions \cite{Randall:1999ee, Randall:1999vf, Lykken:1999nb, Kaloper:1999sm}. In particular, the warped extra-dimensional model due to Randall-Sundrum and its various modifications \cite{Randall:1999ee,Das:2007qn} drew special attention as it hinges crucially on the nature of gravitational interaction at high energies. 

The warped geometry models in general consist of a five dimensional anti de-Sitter (AdS) bulk spacetime within which two 3-branes (four dimensional spacetime) are embedded with $S^1/Z_2$ orbifolding along the extra dimension. The branes are normally assumed to be flat and the separation between the branes is known as the radion or the modular field, whose vacuum expectation value represents the stable distance between the two branes. The radion field, capturing the imprints of higher dimensions in the effective four dimensional dynamics, has several phenomenological and cosmological implications (see for example, \cite{PhysRevLett.84.2080,Davoudiasl:2000wi,Davoudiasl:1999tf,Csaki:1999mp,NIHEI199981,
PhysRevD.61.064003,Csaki:2000zn,Csaki:1999jh,PhysRevD.80.055029, PhysRevD.81.036011,Das:2016dtz, PhysRevD.93.085035,PhysRevD.62.046008,PhysRevD.69.084010,  Banerjee:2017lxi}) and the question of its stabilization is crucial since it gets related to the gauge hierarchy problem in one guise or the other.
 
Several mechanisms have been proposed so far to stabilize the radion field, but most of them have introduced a massive scalar field in the bulk spacetime following \cite{Goldberger:1999uk,GOLDBERGER2000275}. By suitably adjusting the boundary values of the bulk scalar on the two 3-branes the necessary warping for resolution of the gauge hierarchy problem can be achieved without introducing any other hierarchical length scales. However in such contexts the 3-branes are taken to be flat and the origin of the bulk scalar field remains unexplored. This paves way for a  generalized version of the warped geometry models where 3-branes are considered to be non-flat \cite{Das:2007qn} or higher curvature terms are included in the action \cite{Chakraborty:2016gpg,Chakraborty:2016ydo,Chakraborty:2014zya}. It is indeed possible to stabilize the radion field using higher curvature corrections to the Einstein-Hilbert action \cite{Chakraborty:2016gpg,Anand:2014vqa}, however in this work we will concentrate on the consequences of having a non-flat 3-branes on radion stabilization. The non-flat character induces a cosmological constant on the 3-branes resulting into de Sitter or anti-de Sitter branes with distinct phenomenological implications, see e.g., \cite{Das:2007qn,Mitra:2009jw,Chakraborty:2013ipa,Chakraborty:2015zxc,Lipatov:2016ayn,Das:2016dtz,Arun:2016csq,Bazeia:2014dea,Cox:2013rva,Anand:2014vqa,Paul:2018kdq,Das:2017htt}. In a recent work \cite{Banerjee:2017jyk}, it was shown that if the non-flat 3-branes are considered dynamical, then a potential for the radion field is naturally generated, having a turning point in the de-Sitter universe such that stability of radion and gauge hierarchy can be simultaneously addressed. The fact that the radion potential attains a stability for a de Sitter brane-world seems to be consistent with the description of the present accelerated expansion of the universe.

In this work, we consider non-flat branes and study the dynamical evolution of the radion field to its stable value in the context of early universe cosmology. In particular, we would like to explore whether the radion potential appearing due to non-flatness of the 3-branes can result into a sensible inflationary paradigm along with stabilization of the radion field. The early inflationary epoch of our universe is important in many respects, since  it tries to provide an explanation for the horizon and the flatness problems \cite{Fixsen:1993rd, Smoot:1992td, 0067-0049-192-2-18} associated with the hot Big Bang model of our universe \cite{PhysRevD.23.347, LINDE1982431, LINDE1982389, STAROBINSKY198099, PhysRevLett.48.1220, Mukhanov:1981xt}. In inflationary paradigm these problems are overcome by advocating a rapid exponential expansion of the Universe after the Big Bang for a very brief period of time. Within the realm of the above paradigm it is indeed possible to provide a theoretical framework that can produce primordial fluctuations which may later act as the seed for large scale structure formation in the Universe \cite{MUKHANOV1992203, Ade:2013kta, PhysRevD.71.043518, 10.1007/1-4020-3161-0_7}. In normal inflationary scenario, a scalar field known as the inflaton, with a suitable potential is responsible for the exponential expansion and subsequently provide an exit from the rapidly expanding phase as well \cite{Linde:2005ht, Kinney:2003xf, Langlois:2004de, Riotto:2002yw, PhysRevD.50.3746, PhysRevD.57.4614, PhysRevD.51.477, Kumar:2014tya}. In the present context, we would like to portray the radion field, encapsulating the features of extra dimensional physics, as the inflaton field. Further using the radion potential, we would like to explore whether the radion field itself can undergo a slow roll inflation and subsequently achieve a successful exit once the inter-brane separation attains its stable value.

The paper is organized as follows: In \ref{Rad_nFlat_Rev}, we review the generalization of the warped braneworld models to include dynamical, non-flat branes. \ref{Rad_Inf_Evo} is devoted to study the evolution of the radion field to its stable value in the backdrop of the inflationary era and the possibility of the radion field playing the role of inflaton has been explored. Finally, we summarize the main findings of the work and conclude with a discussion of our results in \ref{Conclusion}. Some important calculations associated with the four dimensional effective action have been presented in \ref{appendix_action}.

\textit{Notations and Conventions:} Throughout the paper, the Greek indices have been used to label the four dimensional spacetime coordinates. The metric convention adopted is mostly positive, i.e., (-,+,+,+). Finally, the fundamental constants $c$ and $\hbar$ are set to unity. 
\section{Warped Brane World with non-flat Branes: A Brief Review}
\label{Rad_nFlat_Rev}

The RS warped brane world model \cite{Randall:1999ee} was designed to provide a plausible resolution to the gauge hierarchy problem. It comprises of a 5-dimensional AdS bulk with a negative cosmological constant $\Lambda$ and a single extra dimension inheriting $S^1/Z_2$ orbifold symmetry. In such a scenario our universe is described by a five dimensional metric,
\begin{align}
{ds}^2=e^{-2A(r_c, \phi)}\eta_{\mu \nu} {dx}^{\mu}{dx}^{\nu} + r_c ^2 {d\phi}^2 
\label{Eq1}
\end{align}
where, $\phi$ represents the extra coordinate, $r_c$ the compactification radius and $e^{-2A}$ is the warp factor with $A=kr_c|\phi|$ and $k=\sqrt{-\Lambda/24M^3}$ with $M$ being the five dimensional Planck mass. Two 3-branes are located respectively at the two orbifold fixed points $\phi=0$ (hidden brane) and $\phi=\pi$ (visible brane). As evident from \ref{Eq1}, in the above scenario the two 3-branes are flat, thanks to the exact cancellation between the brane tension and the cosmological constant induced on the brane \cite{PhysRevD.62.024008}. Presence of the warp factor makes it possible to bring down the Higgs mass from the Planck scale ($\approx 10^{19} \textrm{GeV}$) to the TeV scale ($\approx 10^{3} \textrm{GeV}$) on the visible brane. This crucially hinges on the choice of $kr_c$, whose numerical value if taken to be around $\sim 12$, can produce the necessary warping of the Higgs mass without bringing any new energy scale in the theory.

However it is clear that our universe is certainly non-flat as it often comprises of physical situations, e.g., expanding universe, where highly non-flat metrics are invoked. Thus a generalization of the RS model to incorporate non-flat branes is important. It turns out that such a generalization is most naturally achieved by replacing $\eta _{\mu \nu}$ by $g_{\mu \nu}$ in \ref{Eq1}. However in this case the warp factor gets modified and the four dimensional Einstein's equations ensure that the metric $g_{\mu \nu}$ is maximally symmetric, i.e., except for flat spacetime it can be either de-Sitter or anti de-Sitter. Since current observations suggest the spacetime to be de-Sitter, we will mainly consider that option, in which case there will be a non-zero and positive brane cosmological constant $\Omega$, induced from the bulk. In the context of de-Sitter brane, with $\Omega >0$, the warp factor is given by \cite{Das:2007qn}:
\begin{align}
e^{-A}=\omega \sinh\left(\ln\frac{c_2}{\omega}-kr_c |\phi| \right)
\label{Eq2}
\end{align}  
where, $\omega=(\Omega/3k^2)$ a dimensionless constant directly proportional to the brane cosmological constant $\Omega$ and $c_2=1+\sqrt{1+\omega ^2}$ is another constant defined for later convenience. In this case as well for $\Omega \sim 10^{-124}$ (the value of the present day cosmological constant in Planckian units), $kr_c \pi\sim 16\ln10$ can produce the required warping of the Higgs mass on the visible brane as in the RS scenario.

On the other hand, it is also possible to have $\Omega <0$ corresponding to anti de-Sitter brane in which case the warp factor is given by $e^{-A'}=\omega \cosh(\ln\frac{\omega}{c_1}+kr_c |\phi|)$, where $c_1=1+\sqrt{1-\omega ^2}$
\cite{Das:2007qn}. However in what follows we will mainly be interested in the de-Sitter brane. Interestingly introduction of non-flat branes not only induces a positive cosmological constant on the visible brane consistent with the present observed value but also successfully addresses the gauge hierarchy problem by appropriately warping the Higgs mass to the TeV scale.
\subsection{Radion Stabilization in the non-flat Warped Braneworld Scenario}

As in the original RS scenario, in the presence of non-flat branes as well, the resolution of the gauge hierarchy problem is highly sensitive to the stable distance between the two branes, i.e., $r_c$. Therefore, it is more than essential to devise a mechanism to stabilize the inter-brane distance appropriately so that the required warping of the Higgs mass can be achieved. In the context of RS model with flat branes, the stabilization of the inter-brane separation is achieved \cite{Goldberger:1999uk} by the introduction of a bulk scalar field in the bulk action and thereby generating a potential for $r_c$ in the 4-dimensional effective theory. Intriguingly, the minima of the potential corresponds to a particular value of $r_c$ which produces the necessary warping. Further in \cite{GOLDBERGER2000275} the distance between the branes is allowed to vary and as a consequence a new 4-dimensional field $T(x)$ comes into existence in the effective action. This field, known as the radion or the modulus, captures 
the imprints of higher dimensions in the effective 4-dimensional Lagrangian and opens up a new window to search for higher dimensions in the collider experiments. 

However, if the warp factor corresponding to the the scenario involving non-flat branes is considered (as given by \ref{Eq2}, with $r_c$ being replaced by $T(x)$), then it turns out that the radion field generates its own potential and no additional field is required to stabilize the modulus \cite{Banerjee:2017jyk}.
For a derivation of the potential one is referred to \cite{Banerjee:2017jyk} and Appendix A for more details. The nature of the potential depends on the de-Sitter or the anti de-Sitter character of the 3-branes. Considering a de-Sitter brane, i.e., with induced brane cosmological constant $\Omega$ being positive, the radion potential appearing in the four dimensional effective action takes the following form,
\begin{align}
V\left(\frac{\Phi}{f}\right)=6\omega^4 \ln \left(\frac{\Phi}{f}\right)-\frac{3}{2}\omega^2c_2^2\left(\frac{\Phi^2}{f^2}\right)+\frac{3}{2}\omega^2c_2^2+\frac{3}{2}\frac{\omega^6}{c_2^2}\left(\frac{f^2}{\Phi^2}\right)-\frac{3}{2}\frac{\omega^6}{c_2^2} 
\label{Eq3}
\end{align}
Here $\Phi \equiv f\exp\{-kT(x)\pi\}$ and $f=\sqrt{6M^3c_2^2/k}$, with $\omega$ and $c_{2}$ defined earlier. Since instead of $T(x)$, the field $\Phi$ appears in the potential, we will denote $\Phi$ as the modular field which has mass dimensions. 
In the subsequent sections, we will deal with the dimensionless modular field $(\Phi/f)$ for convenience.
It is worthwhile to point out that since $T(x)$ represents the distance between the two branes, it cannot be negative and hence $(\Phi/f)$ must remain within 0 to 1. This will turn out to be a serious constraint while discussing early universe cosmology in the later parts of this work. 

Further, note that in the limit $\omega \rightarrow 0$ the potential term identically vanishes, as it should \cite{GOLDBERGER2000275} and it has an inflection point at $(\Phi_i/f)=\omega/c_2$. Existence of such an inflection point can be easily guessed from the fact that radion potential inherits both convex and concave terms, e.g., both $\Phi^{2}$ and $\Phi^{-2}$ terms appear in the potential. It is well known that existence of inflection point can lead to viable inflationary scenario, which has been studied extensively in the context of particle physics and string-inspired models in recent years \cite{Okada:2017cvy,Okada:2016ssd,Dimopoulos:2017xox,Baumann:2007np,Baumann:2007ah,Allahverdi:2006iq,Allahverdi:2006we,BuenoSanchez:2006rze}. Thus in this case as well, the existence of inflection point will have important consequences in the inflationary scenario. Given the above set of results it is important to explore the evolution of the radion field to its stable value in the backdrop of early universe cosmology, which we will address in the next section.
\section{Inflationary Epoch and Stabilization of the modulus for non-flat branes}
\label{Rad_Inf_Evo}

In the previous section we have observed that the non-flat character of the branes can generate a modular potential which has an inflection point for the de-Sitter branes. It is presumed that the stabilization of the radion field under the potential given by \ref{Eq3} was effective in the very early stages of the universe. This suggests to look for any curious connection between radion stabilization and the inflationary phase of the universe, which also took place at the same energy scale. In particular we would like to explore whether it will be possible for the radion field itself to trigger the inflation, while the exit from the same will result into stabilization of the radion field. Keeping the above in mind, in this section we will try to understand the dynamical evolution of the modulus to its stable value in a general context and investigate if the above can explain the inflationary scenario and hence provide a successful exit from the same.
\subsection{Radion Stabilization: General Analysis}

In this section we will present a general analysis of radion stabilization in the presence of non-flat branes. The results derived here will be used subsequently in the context of early universe cosmology. The bulk action is given by,
\begin{align}
S=S_{gravity}+S_{vis}+S_{hid} \label{Eq4}
\end{align}
where,
\begin{align}
S_{gravity}=\int_{-\infty} ^{\infty} d^4x \int_{-\pi} ^{\pi} d\phi \sqrt{-G}(2M^3 {\bf R} - \Lambda ) \label{Eq5}
\end{align}

\begin{align}
S_{vis}=\int_{-\infty}^{\infty} d^4x \sqrt{-g_{vis}} (L_{vis} -V_{vis}) \label{Eq6}
\end{align}

\begin{align}
S_{hid}=\int_{-\infty}^{\infty} d^4x  \sqrt{-g_{hid}}(L_{hid} -V_{hid})\label{Eq7}
\end{align}
with $\textbf{R}$, the 5-dimensional Ricci scalar. The metric ansatz satisfying the Einstein's equations is taken to be,
\begin{align}
{ds}^2=e^{-2A(x,\phi)}g_{\mu \nu} {dx}^{\mu}{dx}^{\nu} + T(x)^2 {d\phi}^2.  \label{Eq8}
\end{align}
such that the non-flat character of the 3-branes is incorporated, while the warp factor is given by \ref{Eq2}, with $r_c$ replaced by $T(x)$. 
It is customary for the purpose of radion stabilization to start with an effective four-dimensional action, which is obtained by integrating the extra-dimensional coordinate in the bulk action given by \ref{Eq4}. The 4-d effective action thus derived 
can be separated into three parts and is given by,
\begin{align}
~^{(4)}\mathcal{A}_{\rm tot}=~^{(4)}\mathcal{A}_{\rm curv}+~^{(4)}\mathcal{A}_{\rm kinetic} +~^{(4)}\mathcal{A}_{\rm pot} 
\label{Eq9}
\end{align}
Here the first term, namely $~^{(4)}\mathcal{A}_{\rm curv}$ stands for the curvature dependent part of the action. Since the bulk action represents Einstein's gravity, the curvature part of the effective four-dimensional action will be proportional to the 4-dimensional Ricci scalar $R$ and is given by,
\begin{align}
~^{(4)}\mathcal{A}_{\rm curv}=2M^{3}\int d^4 x~\sqrt{-g}~R~
\left\{\frac{c_2 ^2}{4k}+\frac{\omega^2}{k}\ln\left(\frac{\Phi}{f}\right)+\frac{\omega ^4}{4 k c_2^2}\left(\frac{f^2}{\Phi ^2}\right)-\frac{\omega^4}{4kc_2^2}-\frac{c_2 ^2}{4k}\left(\frac{\Phi ^2}{f^2}\right)\right\} 
\label{Eq10}
\end{align} 
The second part corresponds to the kinetic part of the action for the radion field and hence is proportional to $\partial _{\mu}\Phi \partial ^{\mu}\Phi$, which yields,
\begin{align}
~^{(4)}\mathcal{A}_{\rm kinetic}=\int d^4x~\sqrt{-g}~\left(-\frac{1}{2}\partial_\mu \Phi \partial^\mu \Phi\right)
\left\{1+8\frac{M^3}{k}\omega^2\left(\frac{1}{\Phi ^2}\ln\frac{\Phi}{f}\right)-\frac{6M^3}{k}\frac{\omega^4}{c_2 ^2}\left(\frac{f^2}{\Phi ^{4}}\right)\right\} 
\label{Eq11}
\end{align}
and finally we have the potential term, which corresponds to,
\begin{align}
~^{(4)}\mathcal{A}_{\rm pot}=-2M^3 k\int d^4x~\sqrt{-g}~V\left(\frac{\Phi}{f}\right) 
\label{Eq12}
\end{align}
Here $V(\Phi/f)$ is given by \ref{Eq3}. For detailed derivation of the effective action $^{(4)}\mathcal{A}_{\rm tot}$ one is referred to Appendix A. Note that the effective four dimensional action as presented in \ref{Eq9} involves a coupling of the Ricci scalar to the radion field \cite{Karam:2018squ} and hence is in Jordan frame. The coupling to the Ricci scalar will be denoted by $h(\Phi/f)$ and as evident from \ref{Eq10} has the following structure,
\begin{equation}
h(\Phi/f)=\left\{\frac{c_2^2}{4}+\omega^2 \ln\left(\frac{\Phi}{f}\right)+\frac{\omega^4}{4c_2^2}\left(\frac{f^2}{\Phi^2}\right)-\frac{\omega^4}{4c_2^2}-\frac{c_2^2}{4}\left(\frac{\Phi^2}{f^2}\right)\right\}=\frac{1}{6\omega ^{2}}V(\Phi/f)
\label{Eq13}
\end{equation}
The last relation follows from the expression for the potential as presented in \ref{Eq3}. We will use this relation extensively later on. Further, the kinetic term of the radion field, presented in \ref{Eq11}, can be written as $-(1/2)\mathcal{G}(\Phi/f)\partial^\mu\Phi\partial_\mu \Phi$, where $\mathcal{G}(\Phi/f)$ stands for the non-canonical coupling of the scalar field, having the following structure,
\begin{align}
\mathcal{G}(\Phi/f)&=1+8\omega^2\frac{M^3}{k}\frac{1}{\Phi^2}\ln\left(\frac{\Phi}{f}\right)-6\frac{\omega^4}{c_2^2}\frac{M^3}{k}\left(\frac{f^2}{\Phi^4}\right) 
\nonumber 
\\
&=1+\frac{4}{3}\frac{\omega^2}{c_2^2}\left(\frac{f^2}{\Phi^2}\right)\ln\left(\frac{\Phi}{f}\right)-\frac{\omega^4}{c_2^4}\left(\frac{f^4}{\Phi^4}\right)
\label{Eq14}
\end{align}
It is worth emphasizing that as the brane cosmological constant, which is proportional to $\omega$, vanishes, the radion potential $V(\Phi/f)$ also disappears. On the other hand, the non-canonical coupling $\mathcal{G}(\Phi/f)$ becomes unity and hence in this limit the kinetic term of the radion field turns canonical. 

As stated earlier, the introduction of non-flat branes generate a non-trivial potential for the radion field. Thus we have at our disposal a scalar field with a potential, having significant contribution in the near Planckian energy scales. Such energy scales were realized in the context of early universe cosmology and to be consistent with current day observations it has to be dominated by a scalar field with a potential. Thus everything seems to be falling in place, raising the intriguing question, whether the radion field with its potential can mimic the early inflationary stages of our universe? In order to answer this question we transform from the Jordan frame to the Einstein frame such that the coupling of the modulus with the Ricci scalar is removed. The above transformation is achieved by scaling the metric $g_{\mu \nu}$ in the Jordan frame, such that $g_{\mu \nu}\rightarrow \Psi ^{2}(x)g_{\mu \nu}$, where $\Psi(x)$ as of now is an arbitrary scalar function. The above scaling transformation is known as conformal transformation \cite{PhysRevD.65.023521,RevModPhys.82.451,DeFelice2010,PhysRevD.92.026008,0264-9381-14-12-010,BARROW1988515,1475-7516-2005-04-014,PhysRevD.76.084039,1475-7516-2013-10-040} to the metric $g_{\mu\nu}$. Denoting the Einstein frame metric as $\hat{g}_{\mu \nu}$ we obtain: $\hat{g}_{\mu\nu}=\Psi^2 (x)g_{\mu \nu}$ and equivalently $\hat{g}^{\mu\nu}=(1/\Psi^2)g^{\mu \nu}$.

Given the above transformation of the metric tensor from Jordan frame to Einstein frame one can easily compute the Ricci scalar $\hat{R}$ in the Einstein frame which is related to the Ricci scalar $R$ in the Jordan frame by,
\begin{equation}
\hat{R}=\left[\frac{R}{\Psi^2}
-\frac{6}{\Psi^3}g^{\mu\nu}\nabla_\nu\nabla_\mu\Psi \right]
\label{Eq15}
\end{equation}
in four dimensions.
Inverting the above relation it is possible to write down $R$ in terms of $\hat{R}$, such that the action $~^{(4)}\mathcal{A}_{\rm curv}$ presented in \ref{Eq10} becomes,
\begin{equation}
~^{(4)}\mathcal{A}_{\rm curv}=\left(\frac{2M^3}{k}\right)\int d^4x \sqrt{-g}\left[\hat{R}\Psi^2 +\frac{6}{\Psi}\square \Psi\right]h(\Phi/f) \label{Eq16}
\end{equation}
where one may use the fact that $\sqrt{-g}=\Psi^{-4}\sqrt{-\hat{g}}$ to get an action with $\hat{g}_{\mu\nu}$ as the dynamical variable. In order to arrive at the action in the Einstein frame we will have to eliminate any coupling of the scalar field to the Ricci scalar, which demands the following choice for the conformal factor $\Psi$: $\Psi\equiv \sqrt{h(\Phi/f)}$. On the other hand, the second term of \ref{Eq16}, modulo a total divergence term, leads to an additional contribution to the kinetic term. Therefore the kinetic term in the Einstein frame gets modified to, $-(1/2)\mathcal{\hat{G}}(\Phi/f)\partial^\mu\Phi\partial_\mu \Phi$, where the non-canonical coupling of the radion field reads, 
\begin{align}
\hat{\mathcal{G}}(\frac{\Phi}{f})&=\frac{\mathcal{G}(\frac{\Phi}{f})}{h(\frac{\Phi}{f})}+\frac{1}{c_2^2}\bigg[\frac{h^\prime(\Phi/f)}{h(\Phi/f)}\bigg]^2  
\label{Eq17}   
\end{align}
where `prime' denotes differentiation with respect to $\Phi/f$. Finally the potential term after the above conformal transformation to the Einstein frame becomes, $\hat{V}(\Phi/f)$, which is related to the original potential by the following relation: 
\begin{align}
\hat{V}(\Phi/f)=\frac{V(\Phi/f)}{h(\Phi/f)^2}=\frac{6\omega^2}{h(\Phi/f)}
\label{Eq18} 
\end{align}
Finally by incorporating all the ingredients the complete four-dimensional effective action in the Einstein frame is given by,
\begin{align}
~^{(4)}\mathcal{A}_{\rm tot}^{\rm E}=\int d^4 x \sqrt{-\hat{g}}\Bigg[\underbrace{\frac{2M^3}{k}\hat{R}}_{L_{\rm curv}} 
\underbrace{-\frac{1}{2}\hat{\mathcal{G}}(\Phi/f)\partial^\mu\Phi\partial_\mu\Phi}_{L_{\rm kinetic}} 
\underbrace{-2M^3k\hat{V}(\Phi/f)}_{L_{\rm pot}}\Bigg]  
\label{Eq19} 
\end{align}
Here expressions for the non-canonical term $\hat{\mathcal{G}}(\Phi/f)$ in the kinetic part of the Lagrangian and the potential $\hat{V}(\Phi/f)$ in the potential part, has already been presented in \ref{Eq17} and \ref{Eq18} respectively. Variation of the above action with respect to $\hat{g}^{\mu \nu}$, the metric in the Einstein frame, results in the following gravitational field equations,
\begin{align}
\hat{R}_{\mu\nu}-\frac{1}{2}\hat{R}\hat{g}_{\mu\nu}-\frac{k}{4M^3}\Bigg[\mathcal{\hat{G}}(\Phi/f)\partial_\mu\Phi\partial_\nu\Phi-\frac{1}{2}\hat{g}_{\mu\nu}\mathcal{\hat{G}}(\Phi/f)\partial^\alpha\Phi\partial_\alpha\Phi\Bigg]+\frac{k^2}{2}\hat{V}(\Phi/f)\hat{g}_{\mu\nu}=0 \label{Eq20} 
\end{align}
In an identical manner by varying the total action with respect to the radion field $\Phi$, we obtain the following field equation for the radion field,
\begin{align}
\square \Phi +\frac{\hat{\mathcal{G}}'(\Phi/f)}{2f\hat{\mathcal{G}}(\Phi/f)}\nabla _{\mu}\Phi \nabla ^{\mu}\Phi -2M^{3}k \frac{\hat{V}'(\Phi/f)}{f\hat{\mathcal{G}}(\Phi/f)}=0 \label{Eq21}
\end{align}
Later on we will apply the above equations in the context of early universe cosmology. But before delving into those details let us discuss the nature of the potential and the non-canonical kinetic term from a general point of view. \\
First of all it is important to note that the warp factor as presented in \ref{Eq2}, with $r_c$ replaced by $T(x)$ can be written as,
\begin{align}
e^{-A}=\frac{\omega}{2}\left\{\exp \left[\left(\ln\frac{c_2}{\omega}-kT(x) |\phi| \right) \right]-\exp \left[-\left(\ln\frac{c_2}{\omega}-kT(x) |\phi| \right)\right] \right\}
=\frac{c_{2}}{2}\exp(-kT(x)|\phi|)
-\frac{\omega ^{2}}{2c_{2}}\exp(kT(x)|\phi|) \label{Eq22}
\end{align}
Since, the warp factor must always be positive, it immediately follows from the above expression that $(\Phi/f)=\exp\{-kT(x)\pi\}$ can never be less than $(\omega/c_2)$, which is also its inflection point. This further constrains the physically allowed region of the dimensionless modular field $\Phi/f$, such that $\Phi/f$ can vary only between $(\omega/c_2)$ to 1.

\begin{figure*}[h!]
\centering
\subfloat[\label{Fig_1a}]{\includegraphics[scale=0.84]{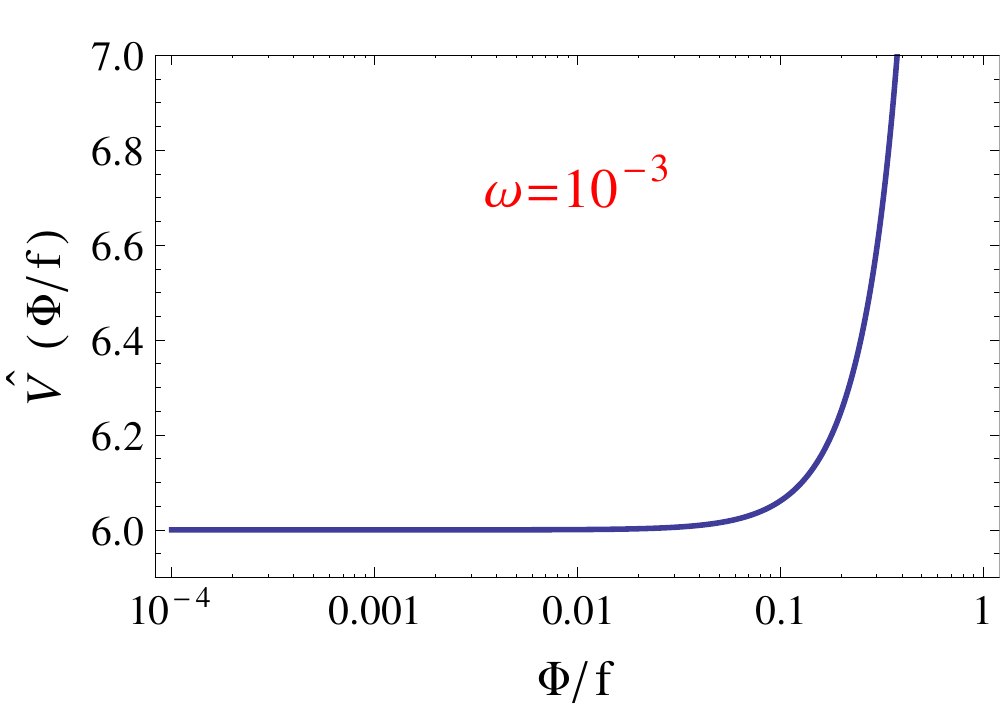}}~~
\subfloat[\label{Fig_1b}]{\includegraphics[scale=0.84]{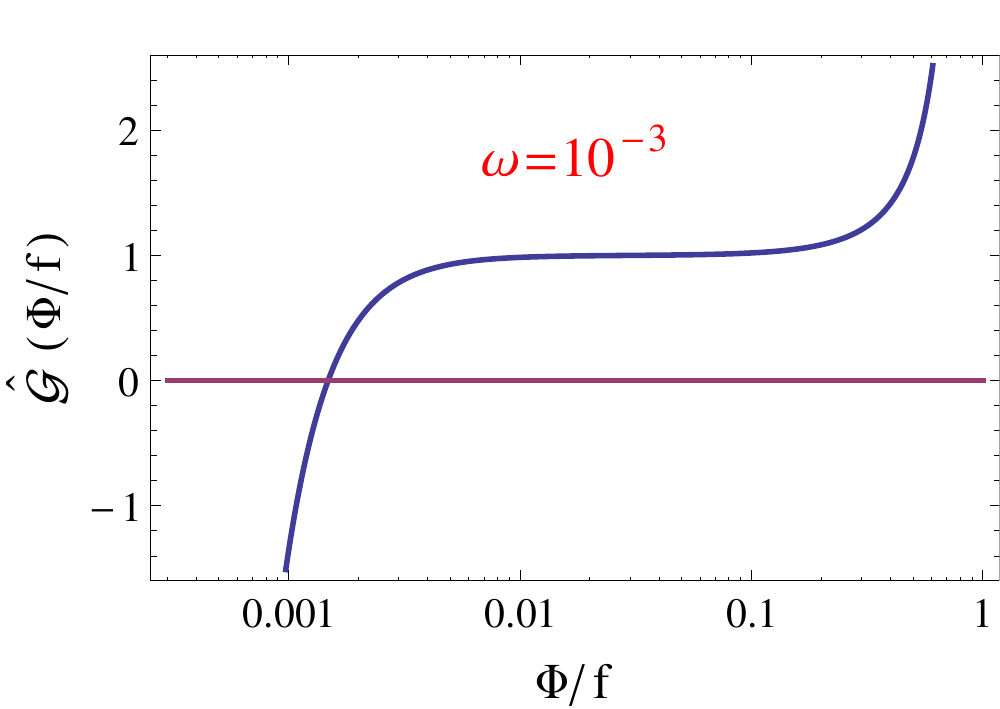}}\\
{\hspace{-1.5cm}\subfloat[\label{Fig_1c}]{\includegraphics[scale=0.92]{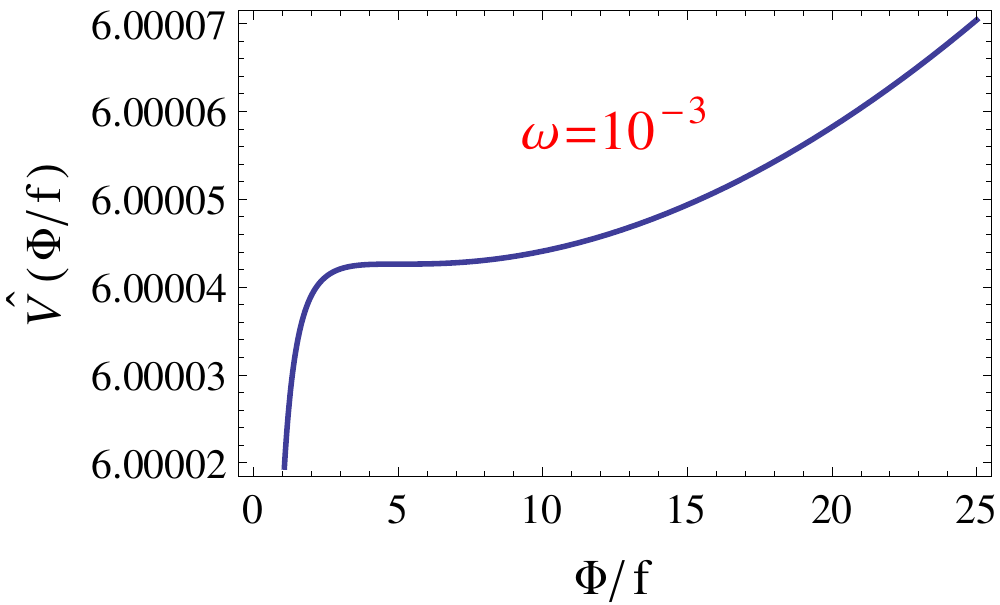}}}~~
\subfloat[\label{Fig_1d}]{\includegraphics[scale=0.82]{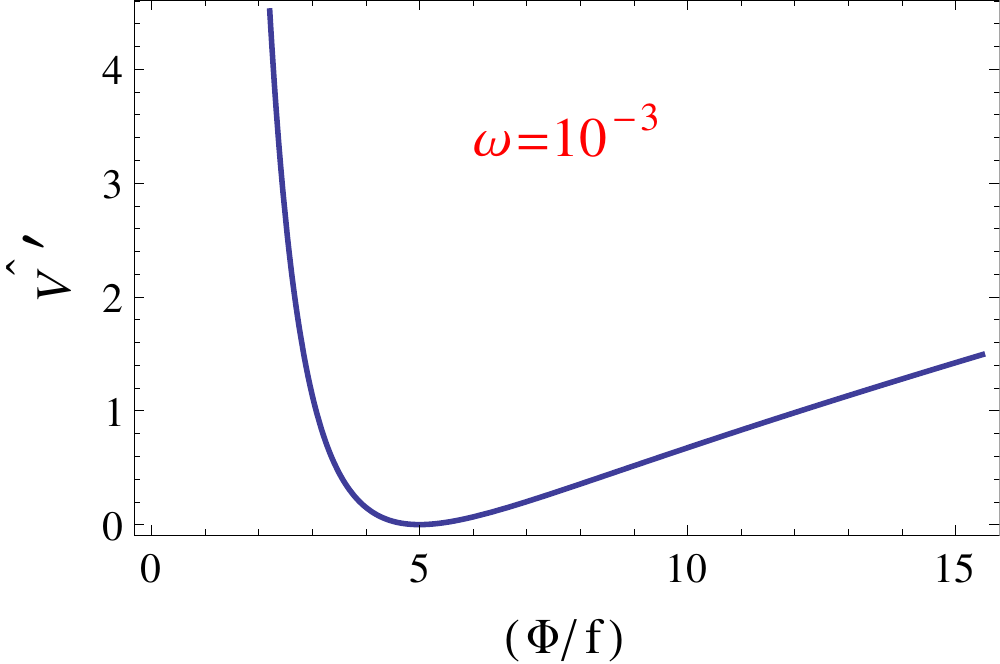}}
\caption{The above figure illustrates the variation of (a) the radion potential $\hat{V}$ (in units of $10^{-6}$) and (b) the non-canonical coupling to the kinetic term $\hat{\mathcal{G}}$, within the allowed range of the radion field $\Phi/f$. Figure 1(c) depicts the nature of the radion potential (in units of $10^{-6}$) near the inflection point $\Phi_i/f=\omega/c_2$ while Figure 1(d) explicitly shows the vanishing of the derivative of $\hat{V}$ ($\hat{V}^\prime$ presented in units of $10^{-8}$) at the inflection point. Note that in Figure 1(c) and Figure 1(d), the $\Phi/f$ in the x-axis is shown in units of $10^{-4}$.
All the above figures are shown for $\omega=10^{-3}$.}
\label{Fig_01}
\end{figure*}

\ref{Fig_1a} and \ref{Fig_1b} illustrates the variation of the radion potential $\hat{V}$ and the behavior of the non-canonical coupling to the kinetic term $\hat{\mathcal{G}}$ with the dimensionless radion field in its allowed range, for $\omega=10^{-3}$. It is intriguing to note that $\hat{\mathcal{G}}$ can be negative for certain values of the allowed range (which implies that the kinetic term can exhibit phantom-like behavior). Interestingly, just as $\Phi_i/f$ is highly sensitive to the value of $\omega$, so is the value of the modulus $\Phi_f/f$ where $\hat{\mathcal{G}}$ changes signature. It is interesting to note that for any $\omega$, 
$\Phi_i/f < \Phi_f/f$ and both scale with the value of $\omega$. For example, for $\omega=10^{-3}$ the value of $\Phi_i/f=\omega/c_2=5\times 10^{-4}$ whereas 
$\Phi_f/f=1.483 \times 10^{-3}$. We will see that this zone will play crucial role in the inflationary era.
The nature of the modulus potential $\hat{V}$ near the turning point $\Phi_i/f=\omega/c_2$ is shown in \ref{Fig_1c} whereas the behavior of the derivative of $\hat{V}$ near the inflection point is explicitly illustrated in \ref{Fig_1d}. It can be demonstrated that the turning point is a point of inflection by computing the higher derivatives of $\hat{V}$ when it turns out that its third derivative is positive at ($\omega/c_2$), while its first and second derivatives vanish at the said point.

\subsection{Implications in Early Universe Cosmology}

Having described the general features of the radion potential and the kinetic term, let us concentrate on its implications for the inflationary paradigm of our universe. In the standard inflationary paradigm a scalar field (often known as the inflaton) with a slow roll potential is invoked to trigger the process of inflation and subsequently the potential also provides a successful exit from the same. Among the multitude of models invoking various potentials for the scalar field to explain inflation, in most of these cases the potentials are introduced by hand and their origin remains unexplained. On the other hand, in this work, the radion field naturally arises in the effective four-dimensional theory and it generates its own potential. Thus, we are tempted to explore if the radion can act as the inflaton field and hence trigger the exponential expansion of the early universe and in the process attain its stable value at the point of exit from the inflationary epoch. 

To study the cosmology of the early universe, it suffices to assume that the four dimensional metric in the Einstein frame $\hat{g}_{\mu\nu}$ is given by the FRW ansatz in the spatially flat form,
\begin{align}
ds^2=-dt^2+a(t)^2\bigg[dx^2+dy^2+dz^2\bigg] \label{Eq23} 
\end{align}
Since the metric depends only on time, the radion field $\Phi$ will have only time dependence without any spatial variation. It is straightforward to determine the dynamical equations governing the evolution of the scale factor $a(t)$ as well as the scalar field $\Phi(t)$ respectively. In particular, the temporal component of the Einstein's equations read,
\begin{align}
3H^2 - \frac{3}{4}c_2^2\mathcal{\hat{G}}(\Phi/f)\bigg(\frac{\dot{\Phi}}{f}\bigg)^2 -\frac{k^2}{2}\hat{V}(\Phi/f)=0  \label{Eq24} 
\end{align}
Along identical lines the spatial components of the Einstein's equations take the following form,
\begin{align}
3H^2 + 2\dot{H} + \frac{3}{4}c_2^2\hat{\mathcal{G}}(\Phi/f)\bigg(\frac{\dot{\Phi}}{f}\bigg)^2  -\frac{k^2}{2} \hat{V}(\Phi/f)=0 \label{Eq25} 
\end{align}
Here, $H=\dot{a}/a$ denotes the Hubble parameter.
In the above homogeneous and isotropic spacetime the field equation satisfied by the radion field $\Phi$ can be immediately written down, having the structure,
\begin{align}
\frac{\ddot{\Phi}}{f}+3H\frac{\dot{\Phi}}{f} +\frac{\hat{\mathcal{G}}^\prime(\Phi/f)}{2\hat{\mathcal{G}}(\Phi/f)}\bigg(\frac{\dot{\Phi}}{f}\bigg)^2 + \frac{k^2}{3c_2^2}\frac{\hat{V}^\prime(\Phi/f)}{\hat{\mathcal{G}}(\Phi/f)} =0 \label{Eq26} 
\end{align}
With all the relevant equations on the desk, let us manipulate them appropriately to extract useful information. For example, subtraction of \ref{Eq24} from \ref{Eq25} results into,
\begin{align}
\dot{H}=-\frac{3}{4}c_2^2\hat{\mathcal{G}}(\Phi/f)\bigg(\frac{\dot{\Phi}}{f}\bigg)^2 
\label{Eq27} 
\end{align}
Similarly, addition of \ref{Eq24} and \ref{Eq25} helps one to get rid of the kinetic term involving $(\dot{\Phi}/f)^{2}$. Using this result and the expression for $\dot{H}$ in \ref{Eq27} we obtain the following equation for the Hubble parameter $H$ as, 
\begin{align}
H^2=\frac{c_2^2}{4}\hat{\mathcal{G}}(\Phi/f)\bigg({\frac{\dot{\Phi}}{f}}\bigg)^2 + \frac{k^2}{6}\hat{V}(\Phi/f) \label{Eq28} 
\end{align}
Having derived the exact expressions for the Hubble parameter and its rate of change with time, we now invoke the slow-roll approximation in which the kinetic term in \ref{Eq28} is considered to be negligibly small in comparison to the potential energy term. 
This gives rise to the following condition,
\begin{align}
\hat{\mathcal{G}}(\Phi/f)\bigg(\frac{\dot{\Phi}}{f}\bigg)^2 << \frac{2}{3}\frac{k^2}{c_2^2}\hat{V}(\Phi/f) \label{Eq29} 
\end{align}
such that the Hubble parameter can have the following simple expression 
\begin{align}
H^2 \approx \frac{k^2}{6}\hat{V}(\Phi/f) \label{Eq30} 
\end{align}
Further, the slow-roll approximation allows us to neglect the $\ddot{\Phi}/f$ and $(\dot{\Phi}/f)^2$ terms present in \ref{Eq26} with respect to the rest of the terms, resulting into,
\begin{align}
\bigg(\frac{\dot{\Phi}}{f}\bigg)\approx -\sqrt{\frac{2}{3}}\frac{k}{3} \frac{\hat{V}^\prime}{\hat{\mathcal{G}}\sqrt{\hat{V}}}\label{Eq31} 
\end{align}

Having derived all the relevant equations we now set to explore whether the radion field can give rise to the inflationary scenario when it is slightly displaced from its inflection point i.e., $\Phi_*/f=\Phi_i/f+\delta$ and provide a successful exit at $\Phi_f/f$ when $\hat{\mathcal{G}}$ crosses zero and becomes positive. Note that the entire duration of inflation takes place in the phantom era when the kinetic term is negative due to the behavior of $\hat{\mathcal{G}}$. Inflation driven by phantom fields have been proposed earlier as well, see e.g., \cite{Piao:2004tq,Piao:2003ty} and thereafter widely studied in \cite{Piao:2007ne,Lidsey:2004xd,GonzalezDiaz:2004df,Capozziello:2005tf,Nojiri:2005pu,Elizalde:2008yf,Baldi:2005gk,Wu:2006wu,Liu:2010dh}.

The epoch of inflation is essentially marked by an accelerated expansion of the scale factor turning into deceleration eventually. Thus, we have $\ddot{a}/a>0$ during the inflationary epoch, while $\ddot{a}/a=0$, marks the exit from the inflationary epoch. The above information is contained within the slow-roll parameter $\epsilon$, which is defined as, 
\begin{align}
\epsilon&=-\frac{\dot{H}}{H^2} \label{Eq32} 
\end{align}
where, $\epsilon$ being less than unity results into the inflationary epoch and $\epsilon \simeq 1$ marks the exit from inflation. Further using \ref{Eq27}, \ref{Eq30} and \ref{Eq31} respectively it is possible to write down the slow-roll parameter in terms of the radion field alone, such that,
\begin{align}
\epsilon \approx \frac{1}{3c_2^2} \frac{1}{\mathcal{\hat{G}}(\Phi/f)}\bigg[\frac{\hat{V}^\prime(\Phi/f)}{\hat{V}(\Phi/f)}\bigg]^2\label{Eq33}
\end{align}
It is quite evident from \ref{Eq31} that $\dot{\Phi}/f$ is positive in the phantom era, diverges when $\mathcal{\hat{G}}$ changes sign and becomes negative when $\mathcal{\hat{G}}$ is positive. This already tells us that during the phantom era $\Phi/f$ increases with time, while outside the phantom epoch $\Phi/f$ decreases with time. This result will have interesting consequences during the exit from the inflationary epoch. 

We next calculate the number of e-foldings of the scale factor over the duration of inflation which is given by,
\begin{align}
N_*=\int_{t_*}^{t_f} H(t)dt  \label{Eq34}
\end{align}
where $t_*$ represents the beginning of inflation and $t_f$ the exit from inflation, such that over the duration of inflation $t_f-t_*$, the modular field evolves from $\Phi_*/f$ to $\Phi_f/f$. Using slow-roll approximation, from \ref{Eq30} and \ref{Eq31}, the expression for the number of e-foldings become,
\begin{align}
N_*=-\frac{3}{2}c_2^2\int_{{\Phi}_{*}/f}^{{\Phi}_{f}/f} d\left(\frac{\Phi}{f}\right) \frac{\mathcal{\hat{G}}(\Phi/f)}{\hat{V}^\prime(\Phi/f)}\hat{V}(\Phi/f) \label{Eq35}
\end{align}
In order to solve the horizon and the flatness problems, which is the prime reason for the introduction of inflation, one should obtain $N_*$ between $50-70$. \\
Since it is evident from the slow-roll approximation that $\epsilon$ diverges when $\mathcal{\hat{G}}$ vanishes, we wish to explore the exact time evolution of the radion field without slow roll approximation. This can be achieved by numerically solving \ref{Eq26} assuming that at the beginning of inflation $t_*$, slow-roll approximation holds. Rather than studying the evolution of $\Phi/f$ with respect to $t$, it is often useful to study the evolution of $\Phi/f$ with respect to the number of e-foldings $N$, which considerably simplifies the calculations. Below we briefly outline the steps:
\begin{itemize}
\item We use the fact that $dN=Hdt$ and define $x \equiv \Phi/f$ for brevity.
\item It can be easily shown that $\frac{dx}{dt}=\dot{x}=H\frac{dx}{dN}=H\underdot{x}$ such that $\ddot{x}=H^2\underddot{x}+H\underdot{H}\underdot{x}$ where $\underdot{x}$ represents derivative of $x$ with respect to $N$.
\item  We rewrite \ref{Eq27} in terms of derivatives with respect to $N$, such that $\underdot{H}=-\frac{3}{4}c_2^2\hat{\mathcal{G}}H\underdot{x}^2$.
\item We substitute $\dot{x}$, $\ddot{x}$ and $\dot{H}$ derived in the last three steps in \ref{Eq26} and divide throughout by $H^2$ that gives us the equation of motion for the scalar field in terms of derivatives with respect to $N$,
\begin{align}
\underddot{x}+3\underdot{x}-\frac{3}{4}c_2^2\mathcal{\hat{G}}(x)\underdot{x}^3+\frac{\mathcal{\hat{G}}^\prime(x)}{2\mathcal{\hat{G}}(x)}\underdot{x}^2+\frac{1}{3c_2^2}\frac{\hat{V}^\prime}{\mathcal{\hat{G}}(x)}\frac{1}{\bar{H}^2}=0 \label{Eq36}
\end{align}
where $\bar{H}=H/k$ is the dimensionless Hubble parameter. 
\item From \ref{Eq28}, it can be shown that $\bar{H}^2=\hat{V}/(6-1.5c_2^2\mathcal{\hat{G}}\underdot{x}^2)$ and knowing the form of $\mathcal{\hat{G}}$ and $\hat{V}$ in terms of $x$ from \ref{Eq17} and \ref{Eq18} respectively, we can solve \ref{Eq36} with appropriate initial conditions.
\end{itemize}

Since it is a second order differential equation we will require two initial conditions to obtain its solution. We fix $x$ and $\dot{x}$ assuming slow-roll condition to be valid at $t_*$ . The choice of $x$ at $t_*$, i.e., $x_*$ depends on $x_i$ which in turn depends on $\omega$. For our model with $\omega=10^{-3}$, $x_*$ is chosen to be $x_i+\delta=\omega/c_2+\delta$ where $\delta=0.095\omega/c_2$ gives the necessary e-folding $\sim 66$. $\underdot{x}_*$ at $t_*$ is obtained from \ref{Eq31}, where $\underdot{x}_*\approx-2 \hat{V}^\prime/(3\mathcal{\hat{G}}\hat{V})$. 

\begin{figure}[h!]
\centering
\subfloat[\label{Fig_2a}]{\includegraphics[scale=0.84]{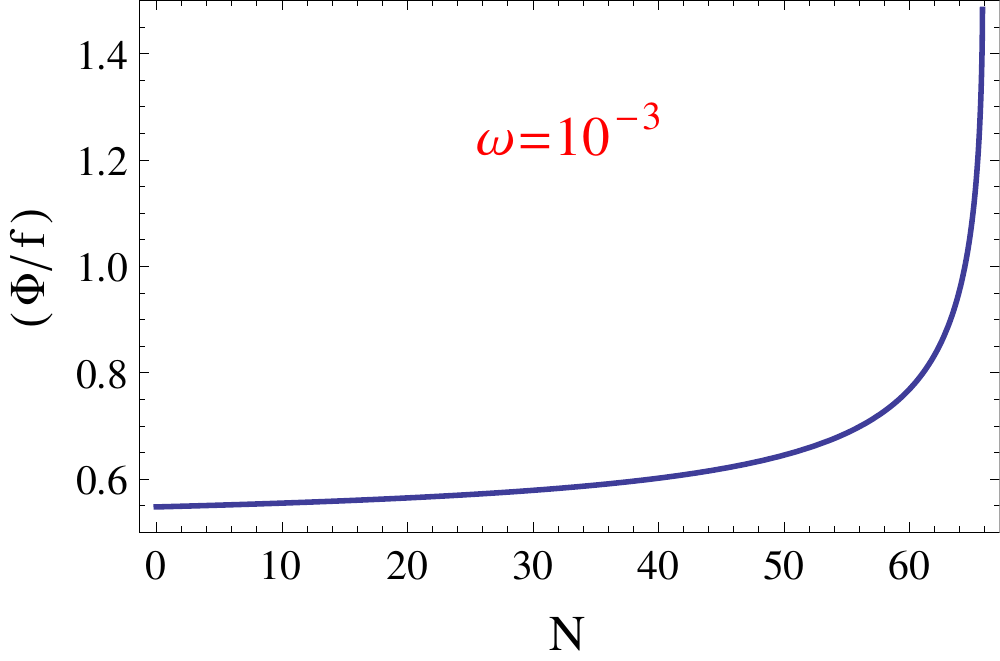}}~~
\subfloat[\label{Fig_2b}]{\includegraphics[scale=0.87]{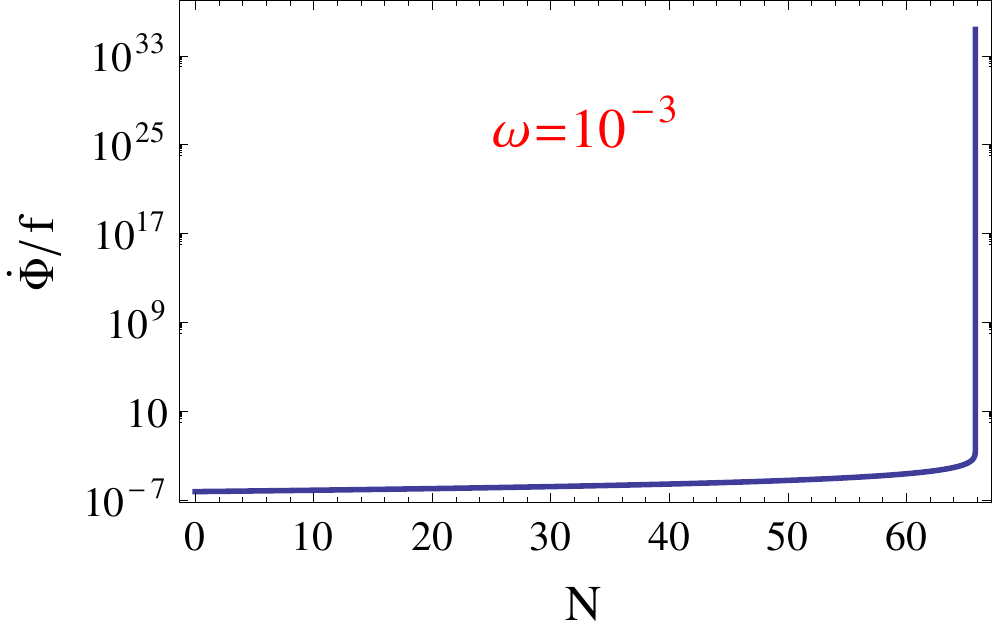}}\\
\caption{The above figure depicts the evolution of (a) the radion field $\Phi/f$ (in units of $10^{-3}$) and (b) the derivative of the radion field $\dot{\Phi}/f$ during the inflationary era. $N$ denotes the number of e-folds. Both the figures are shown for $\omega=10^{-3}$.}
\label{Fig_02}
\end{figure}

Solving \ref{Eq36} gives time evolution of $x$. \ref{Fig_02} represents variation of (a) the dimensionless radion field $x$ and (b) its derivative $\underdot{x}$ with $N$ for $\omega=10^{-3}$.
It indeed turns out that $\Phi_f/f \sim 1.48 \times10^{-3}$ marks the end of inflation such that the number of e-folds achieved is $\sim 66$. This is consistent with the value of $\Phi/f$ where $\mathcal{\hat{G}}$ becomes positive for $\omega=10^{-3}$. The fact that $\underdot{x}$ becomes very large at the end of inflation is evident from \ref{Fig_2b}. This was also approximately predicted from the slow-roll approximation. \\

\begin{figure}[h!]
\centering
\subfloat[\label{Fig_3a}]{\includegraphics[scale=0.84]{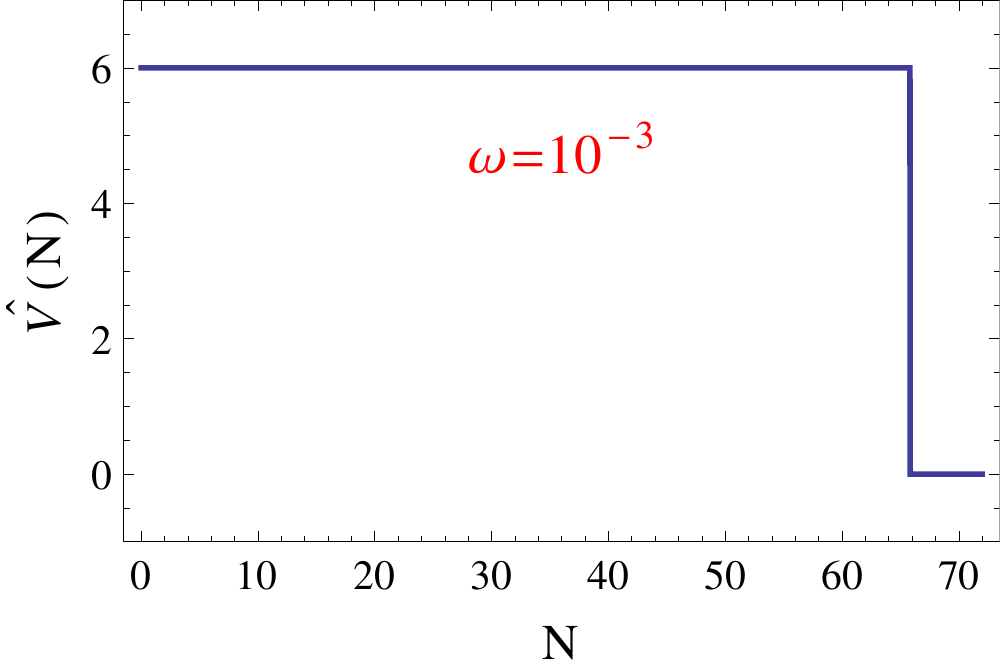}}~~
\subfloat[\label{Fig_3b}]{\includegraphics[scale=0.8695]{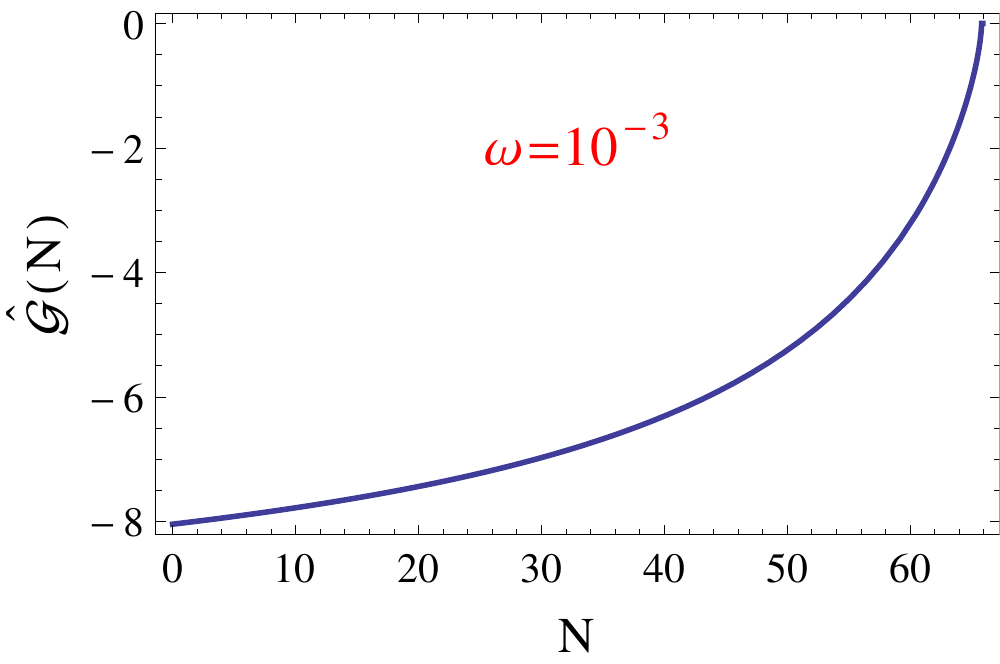}}\\
\caption{The above figure depicts the evolution of (a) the radion potential $\hat{V}$ (in units of $10^{-6}$) and (b) the non-canonical coupling to the kinetic term $\hat{\mathcal{G}}$ during the inflationary era. $N$ denotes the number of e-folds. Both the figures are shown for $\omega=10^{-3}$.}
\label{Fig_03}
\end{figure}

The variation of the potential $\hat{V}$ and the non-canonical coupling $\mathcal{\hat{G}}$ with $N$ is shown in \ref{Fig_3a} and \ref{Fig_3b} respectively. The time evolution of $\hat{V}$ brings out that just at the end of inflation $\hat{V}$ experiences an abrupt jump to a smaller value, while it remains almost constant throughout the inflationary epoch. \ref{Eq30} tells us that the Hubble parameter also remains constant during inflation and scales with $\omega$, i.e., for $\omega=10^{-3}$, $\bar{H}\sim 10^{-3}$. Also \ref{Fig_3b} shows that indeed $\mathcal{\hat{G}}$ crosses zero after $\sim 66$ e-folds, which marks the exit from the phantom era.

From the behavior of the potential in \ref{Fig_3a} it is clear that its derivative diverges at the end of inflation. \ref{Fig_3b} shows that the normal behavior of the kinetic term $\hat{X}=\frac{1}{2}\mathcal{\hat{G}}(\dot{\Phi}/f)^2$ is restored at the point of exit from the inflationary era. With $\mathcal{\hat{G}}$ becoming positive there is a sign reversal in $(\dot{\Phi}/f)$ (see \ref{Eq31}), i.e., now with increase in time $\Phi/f$ would decrease. But the radion field now faces the steep potential barrier which prohibits it to go beyond $\Phi_f/f$. Thus, the radion field can neither attain values greater than $\Phi_f/f$ nor achieve values less than $\Phi_f/f$. This explains the stability of the radion at the point of exit from the inflationary epoch. 

\begin{figure}[h!]
\centering
\includegraphics[scale=0.851]{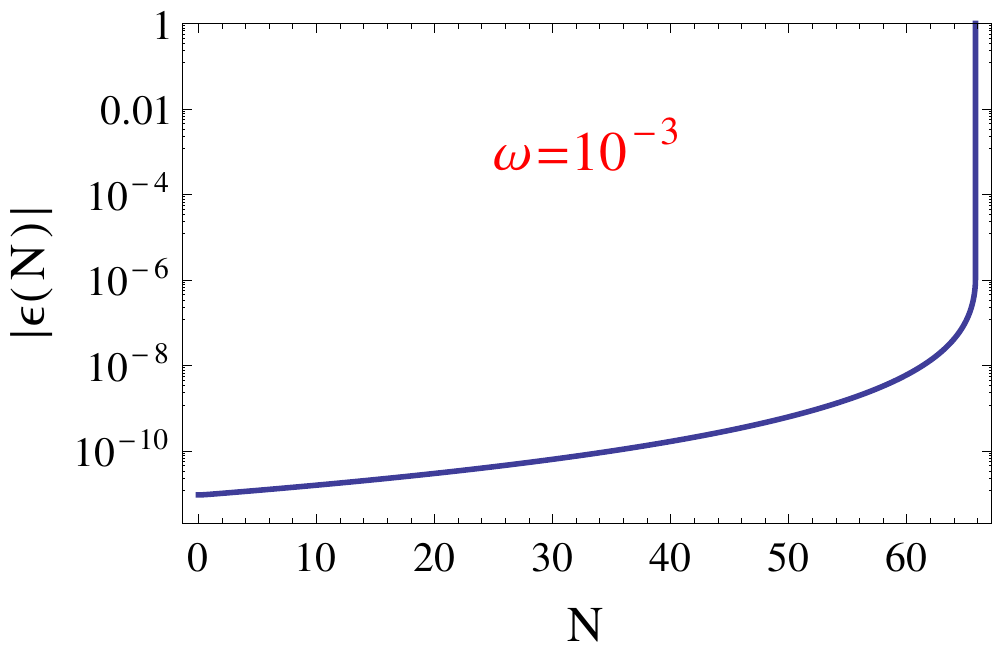}\\
\caption{The above figure shows the evolution of the slow roll parameter $\epsilon$ during the inflationary epoch. $N$ denotes the number of e-folds. The figure is shown for $\omega=10^{-3}$.}
\label{Fig_04}
\end{figure}

In \ref{Fig_04} we show the variation of the slow roll parameter $\epsilon$ with $N$. \ref{Fig_04} shows that indeed slow-roll is a good approximation during inflation. Also note that during much of inflation $\epsilon$ is negative since $\dot{H}$ is positive (see \ref{Eq27}), which in turn implies a very rapid acceleration of the scale factor. This arises due to the non-canonical coupling to the kinetic term, which is negative during the inflationary era. Towards the end when $\mathcal{\hat{G}}$ approaches zero, the acceleration falls drastically such that the slow roll parameter $\epsilon$ attains unity.

Before going into further details, let us briefly discuss the nature of the potential immediately after the end of the inflation. If one is interested in the nature of $\hat{V}$ alone, a Taylor series expansion about the end of inflation, located at $(\Phi_{f}/f)$, will lead to the desired expression, involving both linear and quadratic terms of the potential. However, from the dynamical point of view it is more advantageous to look for the form of Eq.(36) after the end of inflation and ask for the driving potential of $(\Phi/f)$.This can be achieved by noting that around the end of inflation $\underdot{x}$ is large and almost constant (of course, later on it decays, but that is unimportant to this discussion). Thus in Eq.(36), the term depending on $\underdot{x}^{3}$ (which is proportional to $\dot{\Phi}^{3}$) is the dominating contribution. This allows one to write Eq.(36) as $\underddot{x}\simeq -\omega_0^2 x$, or equivalently, $\ddot{\Phi}=-\Omega_{0}^{2}\Phi$ with $\omega _{0}\propto \Omega_{0}$, thus depicting a harmonic oscillator like behaviour around the stable point. Hence the radion field, immediately after the inflation, oscillates around the stabilized value as it moves in an approximated simple harmonic potential.

Another point of interest is to provide an estimate of the mass of the radion field. Since the kinetic term is non-canonical, one has to first make it canonical by means of a field redefinition and then express the potential in terms of the new field and compute its second derivative at $x_f=\Phi_f/f$. In order to accomplish this, we expand $\hat{\mathcal{G}}(x)$ about $x_f$ and retain upto its first order term, i.e.,
\begin{align}
\hat{\mathcal{G}}(x)=\hat{\mathcal{G}}(x_f)+\hat{\mathcal{G}}^\prime(x_f)x +  \mathcal{O}(x^2) +...\label{EqR8}
\end{align}
where $x=\Phi/f$. At $x=x_f$, $\hat{\mathcal{G}}(x)=0$, therefore upto the linear order term near $x=x_f$, $\hat{\mathcal{G}}(x)$ is given by,
\begin{align}
\hat{\mathcal{G}}(x) \approx \hat{\mathcal{G}}^\prime(x_f)x \label{EqR9}
\end{align}
This expansion makes field redefinition much easier and in terms of the new variable $\bar{x}=\bar{\Phi}/f$ we can make the kinetic term canonical, where $\bar{x}$ is given by,
\begin{align}
\bar{x}&=\sqrt{\hat{\mathcal{G}^\prime}(x_f)}\int \sqrt x dx \nonumber \\
&= \sqrt{\hat{\mathcal{G}^\prime}(x_f)}\frac{2}{3}x^{3/2}
\label{EqR10}
\end{align} 
Next, we cast the radion potential $\hat{V}(x)$ in terms of the newly defined field $\bar{x}$ which changes the functional form of the potential, which we denote by $\bar{V}(\bar{x})$. Taylor expanding $\bar{V}(\bar{x})$ about $\bar{x}_f$ gives,
\begin{align}
\bar{V}(\bar{x})&=\bar{V}(\bar{x}_f)+ \frac{d\bar{V}}{d\bar{x}}\bigg|_{\bar{x}=\bar{x}_f}\bar{x}~+~\frac{1}{2} \frac{d^2\bar{V}}{d\bar{x}^2}\bigg|_{\bar{x}=\bar{x}_f}\bar{x}^2 + ...\label{EqR13}
\end{align}
Since the kinetic term is canonical in terms of the field $\bar{x}$, the coefficient of $\bar{x}^2$ in \ref{EqR13} gives an estimate of the mass $m_0$ of the field which turns out to be $m_0\sim 2.193\times10^{-6} k$ (where $k \sim M_{Pl}$). This indicates that the estimated value of the radion mass is consistent with fifth force experiments and cosmology. Note that in \cite{Frolov:2003yi} de-Sitter branes were considered in the context of radion stabilization and that leads to a negative mass for the radion field. While we are getting a positive mass for the radion, as depicted earlier. The difference has to do with the fact that basic premises of these models are distinct. For example, in \cite{Frolov:2003yi} the authors work with the bulk Einstein's equations with a bulk scalar field. This in the spirit of Kaluza-Klein decomposition results into a negative mass squared for the radion field. However, in this work we are never bothered about the bulk dynamics, as we have integrated out the extra dimension and there is no Kaluza-Klein decomposition whatsoever. This being the prime reason behind obtaining a positive radion mass in our analysis.


Further, it is worthwhile to comment on the validity of the effective four dimensional theory in the context of the inflationary scenario. The inflection point, as pointed out earlier, corresponds to the following choice of the scalar field, $(\Phi/f)=(\omega/c_{2})$. During inflation the value of the scalar field was also around the inflection point only. Thus the validity of our effective four dimensional theory, which essentially requires $(\Phi/f)$ to be smaller than the Planck scale, puts some constraint on $\omega$. In particular, we need $\omega\lesssim 10^{-2}$, to safely argue that our effective four dimensional theory remains valid throughout inflation. However, all values of $\omega$ cannot possibly give rise to a viable inflationary scenario and hence it is important to understand whether any fine tuning is required or not. As emphasized earlier, one can arrive at appreciable e-folding and henece a viable inflationary model, as long as $\omega \gtrsim 10^{-16}$. Thus there is a large window with $10^{-16}\lesssim \omega \lesssim 10^{-2}$ and hence there is no appreciable fine tuning involved in $\omega$ space. However, for smaller values of $\omega$, the initial value of the scalar field needs to be chosen close to $(\omega/c_{2})$ and possibly some fine tuning is necessary there.

Finally, in order to be a viable model for inflation, we must estimate the observational quantities associated with the spectrum of primordial fluctuations from our inflationary model. This essentially involves two observables --- (a) the tensor to scalar ratio $r$ and (b) the spectral index of scalar perturbations $n_{s}$. For this purpose it will be advantageous to write down the energy density $\rho_\Phi$ and pressure $p_\Phi$ due to the non-canonical radion field $\Phi/f$ in the spatially flat FRW background. These are respectively given by,
\begin{align}
\rho_{\Phi}=2\hat{X}\frac{\partial L_\Phi}{\partial \hat{X}}-L_\Phi=\hat{X}+\hat{V};\qquad
p_\Phi=L_\Phi=\hat{X}-\hat{V} 
\label{Eq37}
\end{align}
where $\hat{X}$ is the kinetic term for the radion field and $L_\Phi$ represents the Lagrangian due to the radion field. Given the energy density and pressure for the radion field one can immediately compute the ``sound speed" $c_s$ as,
\begin{align}
c_s^2=\frac{p_{\Phi,\hat{X}}}{\rho_{\Phi,\hat{X}}} \label{Eq38}
\end{align}
which is $1$ for our model. The power spectrum for the scalar fluctuations is given by,
\begin{align}
P^\Phi_k=\frac{16}{9}\frac{\rho_\Phi}{\varepsilon_{p}}\frac{1}{c_s(1+\frac{p_\Phi}{\rho_\Phi})} \Bigg|_{aH=c_sk_*} \label{Eq39}
\end{align}
Here $\varepsilon_p=G_N^{-2}$ is related to the four dimensional gravitational constant and the right hand side is evaluated at the time of horizon crossing where $aH=c_sk_*$.
The spectral index for the scalar mode is given by \cite{Garriga:1999hf,Chung:2003iu}
\begin{align}
n_s-1\equiv\frac{d~{\ln}P^\Phi_k}{d~{\ln} k}=-3\bigg(1+\frac{p_\Phi}{\rho_\Phi}\bigg)-\frac{1}{H}\frac{d}{d\Phi}\Bigg[{\ln}\bigg(1+\frac{p_\Phi}{\rho_\Phi}\bigg)\Bigg]\dot{\Phi}-\frac{1}{H}\frac{d}{d\Phi}({\ln}~c_s)\dot{\Phi}+\mathcal{O}(\epsilon ^{2}) 
\label{Eq40}
\end{align}
In the above expression we have kept terms upto lowest order in the slow roll parameters $1+p_\Phi/\rho_\Phi$ and $H^{-1}d{\ln}~c_s/dt$. Note that in the present context the last term in \ref{Eq40} drops out as $c_s=1$. For our model with $\omega=10^{-3}$, it turns out that $n_s \sim 0.967$. Further, in the current scenario the ratio of the tensor fluctuations to the scalar fluctuation is given by \cite{Chung:2003iu},
\begin{align}
r=24c_s\big(1+\frac{p_\Phi}{\rho_\Phi}\big)\bigg|_{t=t_*}=16|\epsilon(t_*)| \label{Eq41}
\end{align}
where, $\epsilon$ is the slow roll parameter given by \ref{Eq32} (once time evolution of $\dot{H}$ and $H^2$ is obtained) and evaluated at the horizon crossing time, $t_*$. It turns out that $r\sim 1.52 \times 10^{-10}$.
In particular, cosmological observations of temperature and polarization anisotropies of the cosmic microwave  background radiation from Planck mission constrains the scalar spectral index to $n_s=0.968 \pm 0.006$ and restricts the tensor-to-scalar ratio $r<0.11$. Joint analysis of results from BICEP2, Keck Array and Planck provides a further tighter constraint on $r<0.09$ \cite{Ade:2015xua,Aghanim:2015xee,Ade:2013kta,MUKHANOV1992203,Lidsey:1995np}. Thus the observables $n_s$ and $r$ derived from this model is consistent with the current observations.

\section{Summary and Discussions}
\label{Conclusion}

In this work we show that warped geometry models with dynamical, non-flat branes can generate a potential for the modulus or radion at the level of four dimensional effective action, without invoking any additional field in the gravitational action. The potential vanishes as the brane cosmological constant $\omega\rightarrow 0$, i.e., in the case of flat branes. The radion potential has a point of inflection located at $(\Phi_{i}/f)=\omega/c_2$ for a non-flat braneworld with positive cosmological constant. This feature holds in both Jordan and Einstein frames, i.e., with or without non-minimal coupling of the radion with Ricci scalar. The kinetic term for the radion field, on the other hand, has a non-canonical coupling which is negative at the inflection point of the radion potential $(\Phi_{i}/f)$ and exhibits zero crossing at a value $(\Phi_{f}/f)$, slightly greater than the inflection point, i.e., $(\Phi_{f}/f)>(\Phi_{i}/f)$ for any $\omega$. In fact, both $(\Phi_{i}/f)$ and $(\Phi_{f}/f)$ scale with $\omega$. The other interesting results we have arrived at corresponds to the following findings:
\begin{itemize}

\item  We have explicitly demonstrated that the radion field can trigger a rapid exponential growth of the early universe, leading to the inflationary paradigm. It turns out that such a rapid growth of the universe takes place in the phantom epoch when $\omega/c_2+\delta<(\Phi/f)<(\Phi_{f}/f)$, where $\delta$ is a small quantity. Moreover, the above approach also provides a successful exit from the inflationary phase as the normal behavior, i.e., positive definiteness of the kinetic term is restored. 

\item We have shown that as the universe exits from inflationary paradigm, the radion field also ceases to be phantom-like. This in turn leads to the desired mechanism responsible for stabilization of the radion field, and is related to the point, where the universe exits from the inflationary epoch. 

\item The scalar spectral index $n_s$ and the tensor-to-scalar ratio $r$ derived from this model are indeed consistent with the recent observations from Planck satellite as well as the BICEP2 and Keck Array. In particular for $\omega \sim 10^{-3}$, the two observables mentioned above turns out to be, $n_s\sim 0.967$ and $r\sim 1.52 \times 10^{-10}$.

\end{itemize}

Although we have demonstrated all of our results for $\omega \sim 10^{-3}$, the radion induced inflation in the phantom era is a generic feature for any value of $\omega\lesssim \mathcal{O}(1)$, such that the zero crossing of the non-canonical kinetic term captured by $\mathcal{\hat{G}}$ is below $(\Phi/f)=1$. We would also like to mention that the value of tensor-to-scalar ratio $r$ estimated from this model is in general quite low for any generic value of $\omega$. The value of $r$ may go upto $\sim 10^{-7}$ on considering higher values of $\omega$, e.g., $\omega\sim \mathcal{O}(1)$. We can also provide a rough estimate of the duration of inflation for this model using the values of $H$ and $\omega$ respectively, which is consistent with conservative estimates. Thus the dynamics of the inter-brane separation, aka radion field, associated with non-flat branes can indeed generate a potential for the same. This in turn can lead to a viable inflationary paradigm for our early universe, while remaining consistent with current observations. Interestingly, stabilization of the radion field can also be achieved. 
\section*{Acknowledgement} 

The research of S.C. is funded by the INSPIRE Faculty Fellowship (Reg. No. DST/INSPIRE/04/2018/000893) from
Department of Science and Technology, Government of India. The research of SSG is supported by the Science and Engineering Research Board-Extra Mural Research Grant No. (EMR/2017/001372), Government of India.
\appendix

\labelformat{section}{Appendix #1}
\labelformat{subsection}{Appendix #1}
\labelformat{subsubsection}{Appendix #1}
\section*{Appendix}

\section{Deriving the four-dimensional effective action and the radion potential}
\label{appendix_action}

In this section we show how the 4-d effective action presented in \ref{Eq9} is derived from the bulk action given in \ref{Eq4}. The bulk action $S$ has three parts,
\begin{align}
S=S_{gravity}+S_{vis}+S_{hid} \tag{A1} \label{Eq42}
\end{align}
where,
\begin{align}
S_{gravity}=\int_{-\infty} ^{\infty} d^4x \int_{-\pi} ^{\pi} d\phi \sqrt{-G}(2M^3 {\bf R} - \Lambda ) \tag{A2} \label{Eq43}
\end{align}

\begin{align}
S_{vis}=\int_{-\infty}^{\infty} d^4x \sqrt{-g_{vis}} (L_{vis} -V_{vis}) \tag{A3}\label{Eq44}
\end{align}

\begin{align}
S_{hid}=\int_{-\infty}^{\infty} d^4x  \sqrt{-g_{hid}}(L_{hid} -V_{hid}) \tag{A4}\label{Eq45}
\end{align}
with $\textbf{R}$, the 5-dimensional Ricci scalar. We assume that the metric ansatz which satisfies the Einstein's equations takes the form,
\begin{align}
{ds}^2=e^{-2A(x,\phi)}g_{\mu \nu} {dx}^{\mu}{dx}^{\nu} + T(x)^2 {d\phi}^2.  \tag{A5} \label{Eq46}
\end{align}
Given the form of \ref{Eq46} it can be shown that the 5-d Ricci scalar $\bf{R}$ assumes the form, 
\begin{align}
{\bf{ R}}&=e^{2A} R + 8\frac{A^{\prime\prime}}{T(x)^2}-20\frac{{A^{\prime}}^2}{T(x)^2}-2e^{2A}\frac{(T(x)_,^{~a})_{;a}}{T(x)} +6e^{2A}(A_,^{~k})_{;k} \nonumber \\ &
-6e^{2A} A_,^{~p} A_{,p}+4e^{2A}\frac{T(x)_{,a}A^{~a}}{T(x)} \tag{A6} \label{Eq47}
\end{align}
Here, $R$ is the 4-d Ricci scalar and ${A^{\prime}}$ represents derivative of $A$ w.r.t the extra-coordinate $\phi$. \\ We consider the branes to mimic a de-Sitter universe such that the form of the warp factor is given by \cite{Das:2007qn},
\begin{align}
e^{-A}=\omega \sinh\left(\ln\frac{c_2}{\omega}-kT(x) |\phi| \right) \tag{A7}\label{Eq48}
\end{align}  
In order to obtain the 4-d effective action $^{(4)}\mathcal{A}_{\rm tot}$, we perform the $\phi$ integration in \ref{Eq43}, \ref{Eq44} and  \ref{Eq45}. Note that $G$ represents the determinant of the 5-d metric given by \ref{Eq46} such that 
$\sqrt{-G}=e^{-4A}T(x)\sqrt{-g}$. \\
In order to derive the curvature part of the effective 4-d action $^{(4)}\mathcal{A}_{\rm curv}$, we compute,
\begin{align}
&\int_{-\infty} ^{\infty} d^4x \int_{-\pi} ^{\pi} d\phi~ \sqrt{-G} ~2M^3 e^{2A}R = 2M^3 \int_{-\infty} ^{\infty} d^4x \int_{-\pi} ^{\pi} ~d\phi \sqrt{-g}T(x)e^{-2A}R   \nonumber \\
=&\frac{2M^3}{k} \int_{-\infty} ^{\infty} d^4x \sqrt{-g}R \Big[\frac{c_2^2}{4}\big(1-\frac{\Phi^2}{f^2}\big)- \frac{\omega^4}{4c_2^2}\big(1-\frac{f^2}{\Phi^2}\big)+\omega^2 ln\frac{\Phi}{f} \Big] \tag{A8}
\label{Eq49}
\end{align}
Note that in the limit $\omega\rightarrow 0$, the last two terms of \ref{Eq49} drop out and we retrieve the familiar result of \cite{GOLDBERGER2000275}.\\ \\
In order to derive the kinetic term $^{(4)}\mathcal{A}_{\rm kinetic}$, we have to consider the last four terms of \ref{Eq47} and integrate over $\phi$.
It can be shown that the fourth term and the last term of \ref{Eq47} mutually cancel except for a surface term which vanishes upon integration over all space. Hence, fifth and sixth terms of \ref{Eq47} actually contribute to $^{(4)}\mathcal{A}_{\rm kinetic}$. \\
It can be shown that apart from a total divergence term the fifth and sixth terms of \ref{Eq47} is given by,
\begin{align}
&2M^3\int_{-\infty}^{\infty} d^{4}x~\int_{-\pi}^{\pi}~d\phi \sqrt{-g} e^{-2A}\Big\lbrace 6T(x)A_{,k}A_{,}^{k}-6 A_{,}^{k}T(x)_{,k}\Big\rbrace  \nonumber \\
=&12M^3\int_{-\infty}^{\infty} d^{4}x \sqrt{-g}~\int_{-\pi}^{\pi}~d\phi~ e^{-2A}\Big\lbrace T(x)k^2|\phi|^2 coth^2\big(ln\frac{c_2}{\omega}-kT(x)|\phi|\big)T(x)_{,k}T(x)_{,}^{k} \Big\rbrace  \nonumber \\ 
-&12M^3\int_{-\infty}^{\infty} d^{4}x \sqrt{-g}~\int_{-\pi}^{\pi}~d\phi~ e^{-2A}\Big\lbrace k|\phi| coth\big(ln\frac{c_2}{\omega}-kT(x)|\phi|\big)T(x)_{,k}T(x)_{,}^{k} \Big\rbrace \nonumber \\
=&\int_{-\infty}^{\infty} d^{4}x \sqrt{-g}\Bigg\lbrace -\frac{6M^3c_2^2}{k}\frac{1}{2} \partial_\mu\frac{\Phi}{f}\partial^\mu\frac{\Phi}{f} + \frac{6M^3}{k}\frac{\omega^4}{c_2^2}\frac{f^4}{\Phi^4}\frac{1}{2}\partial_\mu\frac{\Phi}{f}\partial^\mu\frac{\Phi}{f}-\frac{8M^3}{k}\omega^2 ln\frac{\Phi}{f} \frac{f^2}{\Phi^2}\frac{1}{2}\partial_\mu\frac{\Phi}{f}\partial^\mu\frac{\Phi}{f}\Bigg\rbrace \nonumber \\
=&-\int_{-\infty}^{\infty} d^{4}x \sqrt{-g} \frac{1}{2} \partial_\mu \Phi\partial^\mu\Phi \Bigg\lbrace 1-\frac{\omega^4}{c_2^4}\frac{f^4}{\Phi^4}+\frac{4}{3}\frac{\omega^2}{c_2^2}\frac{f^2}{\Phi^2}ln\frac{\Phi}{f} \Bigg\rbrace \tag{A9}
\label{Eq50}
\end{align}
where, $\Phi \equiv f\exp\{-kT(x)\pi\}$ and $f=\sqrt{6M^3c_2^2/k}$. Again, in the limit $\omega\rightarrow 0$, the kinetic term turns canonical maintaining agreement with \cite{GOLDBERGER2000275}.\\ \\ 
The potential term $^{(4)}\mathcal{A}_{\rm pot}$ has contributions from the second and third terms of \ref{Eq47}, the brane cosmological constant $\Lambda$, $S_{vis}$ and $S_{hid}$. Further the second term of \ref{Eq47} has contributions from the bulk and also from the brane boundary. It can be shown that in \cite{GOLDBERGER2000275} or in the RS scenario all the aforementioned terms mutually cancel and hence there is no potential term.
For the non-flat branes with $\Omega>0$, $V_{vis}$ and $V_{hid}$ are given by \cite{Das:2007qn},
\begin{align}
V_{vis}=24M^3k\Bigg[\frac{\frac{\omega^2}{c_2^2}e^{2kT(x)\pi}+1}{\frac{\omega^2}{c_2^2}e^{2kT(x)\pi}-1}\Bigg] ~~~~~V_{hid}=24M^3k\Bigg[\frac{1+\frac{\omega^2}{c_2^2}}{1-\frac{\omega^2}{c_2^2}}\Bigg] \tag{A10} \label{Eq51}
\end{align}
Adding all the contributions we get,
\begin{align}
~^{(4)}\mathcal{A}_{\rm pot}&=-2M^3k \int_{-\infty}^{\infty} d^{4}x \sqrt{-g} \Bigg\lbrace \frac{c_2^4}{4}\Bigg(1-\frac{\Phi^4}{f^4}\Bigg)-\frac{\omega^8}{4c_2^4}\Bigg(1-\frac{f^4}{\Phi^4}\Bigg)-6\omega^4 ln\frac{\Phi}{f}-2\omega^2c_2^2\Bigg(1-\frac{\Phi^2}{f^2}\Bigg)+2\frac{\omega^6}{c_2^2}\Bigg(1-\frac{f^2}{\Phi^2}\Bigg)\Bigg\rbrace \nonumber \\
&-2M^3k \int_{-\infty}^{\infty} d^{4}x \sqrt{-g} \Bigg\lbrace \frac{3c_2^4}{8}        \Bigg(1-\frac{\Phi^4}{f^4}\Bigg)-\frac{3}{8}\frac{\omega^8}{c_2^4}\Bigg(1-\frac{f^4}{\Phi^4}\Bigg)+ 3\omega^4 ln\frac{\Phi}{f} \Bigg\rbrace  \nonumber \\
&-2M^3k \int_{-\infty}^{\infty} d^{4}x \sqrt{-g} \Bigg\lbrace -c_2^4+\frac{\omega^8}{c_2^4} + 2\omega^2c_2^2 - 2\frac{\omega^6}{c_2^2} \Bigg\rbrace -2M^3k \int_{-\infty}^{\infty} d^{4}x \sqrt{-g} \Bigg\lbrace -\frac{\omega^8}{c_2^4}\frac{f^4}{\Phi^4} +c_2^4\frac{\Phi^4}{f^4}+2\frac{\omega^6}{c_2^2}\frac{f^2}{\Phi^2} \nonumber \\
&-2\omega^2c_2^2\frac{\Phi^2}{f^2} \Bigg\rbrace -2M^3k \int_{-\infty}^{\infty} d^{4}x \sqrt{-g} \Bigg\lbrace -\frac{3c_2^4}{8}\Bigg(1-\frac{\Phi^4}{f^4}\Bigg)+\frac{3}{8}\frac{\omega^8}{c_2^4}\Bigg(1-\frac{f^4}{\Phi^4}\Bigg)+9\omega^4 ln\frac{\Phi}{f}+3\omega^2c_2^2\Bigg(1-\frac{\Phi^2}{f^2}\Bigg)\nonumber \\ 
&-3\frac{\omega^6}{c_2^2}\Bigg(1-\frac{f^2}{\Phi^2}\Bigg)\Bigg\rbrace -2M^3k \int_{-\infty}^{\infty} d^{4}x \sqrt{-g} \Bigg\lbrace \frac{3}{4}\frac{\omega^8}{c_2^4}\frac{f^4}{\Phi^4}-\frac{3}{4}c_2^4\frac{\Phi^4}{f^4}-\frac{3}{2}\frac{\omega^6}{c_2^2}\frac{f^2}{\Phi^2}+\frac{3}{2}\omega^2c_2^2\frac{\Phi^2}{f^2} \Bigg\rbrace \nonumber \\ 
& -2M^3k \int_{-\infty}^{\infty} d^{4}x \sqrt{-g} \Bigg\lbrace \frac{3}{4}c_2^4-\frac{3}{4}\frac{\omega^8}{c_2^4}-\frac{3}{2}\omega^2c_2^2+\frac{3}{2}\frac{\omega^6}{c_2^2}\Bigg\rbrace \label{Eq52} \tag{A11}
\end{align}
This can be simplified to give the final expression for the potential term,
\begin{align}
~^{(4)}\mathcal{A}_{\rm pot}&=-2M^3k \int_{-\infty}^{\infty} d^{4}x \sqrt{-g} \Bigg\lbrace 6\omega^4ln\frac{\Phi}{f}-\frac{3}{2}\omega^2c_2^2\frac{\Phi^2}{f^2} + \frac{3}{2}\omega^2c_2^2    +\frac{3}{2}\frac{\omega^6}{c_2^2}\frac{f^2}{\Phi^2} - \frac{3}{2}\frac{\omega^6}{c_2^2} \Bigg\rbrace  \label{Eq53} \tag{A12}
\end{align}
Note that the potential presented in \cite{Banerjee:2017jyk} leads to \ref{Eq53} after simplification.
In the limit $\omega\rightarrow 0$, (i.e when the branes are flat) the potential term identically vanishes as expected.

\bibliography{radioninflation}
\bibliographystyle{./utphys1}
\end{document}